# Pressure-induced high-temperature superconductivity retained at ambient

Liangzi Deng[1,*], Trevor Bontke[1], Rabin Dahal[1], Yu Xie[2], Bin Gao[3], Xue Li[2], Ketao Yin[4], Melissa Gooch[1], Donald Rolston[1], Tong Chen[3], Zheng Wu[1], Yanming Ma[2], Pengcheng Dai[3], and Ching-Wu Chu[1,5,*]

[1]Department of Physics and Texas Center for Superconductivity, University of Houston, Houston, TX

[2]International Center for Computational Method and Software and State Key Laboratory of Superhard Materials, College of Physics, Jilin University, Changchun, China

[3]Department of Physics and Astronomy, Rice University, Houston, TX

[4]School of Physics and Electronic Engineering, Linyi University, Linyi, Shandong, China

[5]Lawrence Berkeley National Laboratory, Berkeley, CA

*Corresponding authors: ldeng2@central.uh.edu and cwchu@uh.edu

To raise the superconducting-transition temperature ($T_c$) has been the driving force for the long, sustained effort in superconductivity research. Recent progress in hydrides[1-3] with $T_c$s up to 287 K under 267 GPa has heralded a new era of room-temperature superconductivity (RTS) with immense technological promise. Indeed, RTS has lifted the temperature barrier for the ubiquitous application of superconductivity. Unfortunately, formidable pressure is required to attain such high $T_c$s. The most effective relief to this impasse is to remove the pressure needed while retaining the pressure-induced $T_c$ without pressure. Here we show such a possibility in the pure and doped high-temperature superconductor (HTS) FeSe by retaining, at ambient via pressure-quenching (PQ), its $T_c$ up to 37 K (quadrupling that of a pristine FeSe[4]) and other pressure-induced phases[5]. We have also observed that some phases remain stable without pressure at up to 300 K and for at least 7 days. The observations are in qualitative agreement with our *ab initio* simulations using the solid-state nudged elastic band (SSNEB) method[6]. We strongly believe that the PQ technique developed here can be adapted to the RTS hydrides and other materials[7-11] of value with minimal effort.

The vast impact of RTS on humanity is limited only by the imagination. Recent reports show that RTS is indeed within reach, although only under high pressure (HP). For instance, $T_c$s above 200 K have been reported in unstable molecular solids (hydrides), i.e., up to 203 K in $H_3S$ under 155 GPa[1], up to 260 K in $LaH_{10}$ under 190 GPa[2], and up to 287 K in C-H-S under 267 GPa[3]; earlier, $T_c$ up to 164 K was reported in the stable cuprate HTS $HgBa_2Ca_2Cu_3O_{8+\delta}$ under 31 GPa[7,8]. The challenge now shifts from raising the $T_c$ to lowering, and better yet removing, the applied pressure ($P_A$) required. Retaining the pressure-induced high-$T_c$ superconducting (SC) phase without pressure will effectively meet this challenge.

It has been shown that most of the alloys used in industrial applications are actually metastable at room temperature and atmospheric pressure[12]. These metastable phases, such as diamond and black phosphorus, possess desired and/or enhanced properties that their stable counterparts lack. They are metastable because they are kinetically stable but thermodynamically not, protected only by an energy barrier (Extended Data Fig. 1). By taking advantage of such energy barriers, lattice and/or electronic, one may therefore be able to stabilize the metastable phase or the "supercool" state at atmospheric pressure via rapid pressure- and/or temperature-quenching. The energy barrier may be fortified by chemical doping; a proper thermodynamic path; and introduction of strains, defects, or pressure inhomogeneity[13]. The pressure-enhanced or -induced SC phase with a high $T_c$ may be considered metastable or supercool and may be stabilized. Last year we retained two SC phases with $T_c$s ~ 3.4 K and 4.0 K from 9 and 30 GPa, respectively, in non-SC elemental Sb[14]. Here we report the first successful retention of pressure-enhanced and -induced SC phases at ambient in the Fe-based HTSs via PQ at a chosen quench-pressure ($P_Q$) and quench-temperature ($T_Q$), i.e., a pressure-enhanced SC phase with a $T_c$ up to 37 K at $P_Q$ = 4.15 GPa and $T_Q$ = 4.2 K in the SC FeSe and a pressure-induced SC phase with a $T_c$ up to 26.5 K at $P_Q$ = 6.32 GPa and $T_Q$ = 4.2 K in the non-SC Cu-doped FeSe. We have also successfully retained the insulating phase induced by pressure above ~ 9 GPa in both samples via PQ. The pressure-quenched (PQed) high-$T_c$ phases have also been found to be stable up to ~ 200 K and for at least 7 days. Our observations have thus demonstrated that the pressure-enhanced or -induced high-$T_c$ phases in HTSs can be retained at ambient via PQ at a chosen $P_Q$ and $T_Q$, suggesting a possible realistic path to the ubiquitous applications of the recently reported RTS.

In the present study, we have chosen single crystals of the SC FeSe[4] and the non-SC Cu-doped FeSe[15] as model HTSs due to their simple structure and chemistry, as well as their large $T_c$ variation under pressure[5,16] and their important role in unraveling HTS[17-19]. Their normalized resistance at 300 K as a function of pressure [$R(P_A)/R(0)$] during pressure-increasing and -decreasing is displayed in Extended Data Fig. 2, which shows a clear hysteresis, suggesting that PQ may be possible. Preliminary boundaries of the orthorhombic (O) – tetragonal (T) – hexagonal (H) phase transitions of FeSe previously reported[16,20] are also shown for later discussion. The T-O transition is suppressed from ~ 90 K at ambient to below 4.2 K at ~ 2 GPa, as indicated by the dashed line at left in the same figure. At ambient, R(T) of FeSe shows a sharp SC transition at 9 K (Extended Data Fig. 3a). The transition broadens under pressure, so the $T_c(P)$ cited hereafter refers to the onset temperature as defined. Figure 1 (blue squares) displays the $T_c$-variation of FeSe with $P_A$: it increases slowly from ~ 9 K at 0 GPa to ~ 15 K below 1.9 GPa; suddenly jumps to ~ 32 K at 1.9 GPa, coinciding with the O-T transition; continues to rise with a broad peak at ~ 40 K around ~ 4 GPa; but finally becomes insulating above ~ 8 GPa as the H phase sets in.

To retain at ambient the above pressure-enhanced $T_c$ of FeSe, we have developed a technique to PQ the sample at different $P_Q$s and $T_Q$s by rapidly removing the $P_A$, under which a desired $T_c$ has been first attained, from the sample in the diamond anvil cell (DAC), as shown in Figs. 2a-f. The temperature-dependent resistance of FeSe at different $P_A$s normalized to those at 70 K, $R(T,P_A)/R(70 K,P_A)$, near the superconducting transitions are exemplified in Figs. 2a-b for $P_A$s = 4.15 GPa *(close to maximum $T_c$ ~ 40 K in the tetragonal phase),* and 11.27 GPa *(non-SC in the hexagonal phase)*, respectively. By following different thermal and pressure protocols as specified in the captions, they demonstrate the generation or destruction of the HP SC phase at $P_A$ (blue), the retention at ambient of the PQed (at 4.2 K) HP SC phase (red), and the thermal annealing effect up to 300 K on the PQed (at 4.2 K) HP phase to ascertain its retention (orange), all carried out sequentially.

As is evident from Fig. 2a, the $T_c$ of the FeSe sample has been enhanced from 9 K at ambient to ~ 39 K under ~ 4.15 GPa (blue). After PQ at 4.15 GPa and 4.2 K, a SC transition with a $T_c$ ~ 37 K is detected at ambient (red). To show that the 37 K-$T_c$ is indeed attained by PQ, we heated the sample up to 300 K before cooling it back down to 4.2 K and found that the PQed SC transition at 37 K is annealed away and replaced by its pre-PQed one, although at a higher $T_c \sim 20$ K (orange) rather than 9 K, presumably because of an unknown irreversible residual strain effect in the sample[21]. Figure 2b shows that FeSe at 11.27 GPa displays a non-SC transition as expected (blue), as does the PQed sample (red). However, the sample regains its SC transition with a $T_c \sim$ 20 K after the PQed phase is annealed off after being heating up to 300 K (orange). To demonstrate the metastability of the PQed SC phases, the SC transition PQed at $P_Q = 4.13$ GPa and $T_Q = 4.2$ K upon sequential thermal cycling to higher temperatures is shown in Fig. 2c. The transition is smoothly shifted downward and becomes sharper (due to the reduced fluctuations at lower temperature). The sudden downward shift in the overall SC transition by $\sim 10$ K after heating up to $\sim 200$ K implies that the PQed phase transforms to the pre-PQed FeSe phase (with strain) and is stable up to 200 K. All $T_c$s of the PQed phases examined at different $P_Q$s and $T_Q = 4.2$ K are summarized in Fig. 1 (red circles).

As mentioned earlier, the PQed phase is metastable, and thus should depend on $P_Q$ and $T_Q$ and detailed electronic and phonon energy spectra of the materials examined. We have therefore repeated the PQ experiments on FeSe by raising only the $T_Q$ to 77 K (Fig. 2d-f). Figure 2d shows that the $T_c$ of FeSe before PQ has been enhanced to $\sim 37$ K at 5.22 GPa (blue); upon PQ, a $T_c \sim$ 24 K is retained at ambient (green) in contrast to the 37 K when $T_Q = 4.2$ K, as shown in Fig. 2a; and the transition returns to $\sim 14$ K on cooling after warming to 300 K, showing that the 24 K transition is associated with the PQed phase. Figure 2e shows that FeSe becomes insulating at 11.12 GPa (blue); the phase is retained at ambient by PQ (green); and the PQed phase remains after heating to 300 K, suggesting that this PQed non-SC phase is stable up to 300 K. The effect of systematic thermal cycling with increasing temperatures on the PQed phase at $P_Q$ =11.12 GPa and $T_Q$ = 77 K is shown in Fig. 2f. All $T_c$s of the PQ phases examined at different $P_Q$s and $T_Q$ = 77 K are also summarized in Fig. 1 (green diamonds). They are all lower than those quenched at $T_Q$ = 4.2 K in general agreement with the competition between the instability of the SC state and thermal excitation.

To avoid possible confusion of the SC state of the pristine FeSe with that of the SC PQed FeSe, we have investigated the Cu-doped FeSe [$(Fe_{0.98}Cu_{0.03})Se$], which is not SC above 1.2 K in its uncompressed state[5,15], as shown in Extended Data Fig. 3b. Under pressure [Fig. 3 (blue squares)], it abruptly becomes SC with a $T_c$ ~ 20 K at 3.11 GPa (inset, Extended Data Fig. 3b); $T_c$ continues to increase with increasing $P_A$ and peaks at ~ 27 K under 6.23 GPa; and at 9.65 GPa, only trace superconductivity was detected down to 1.2 K. Following the same protocols as those for the pure FeSe, we performed PQ on Cu-doped FeSe at different $P_Q$s and $T_Q$s, as shown in Figs. 4a-f. Two examples of $R(T,P_A)$s/$R(50 K, P_A)$ for Cu-doped FeSe are given in Fig. 4a for $P_A$s = 6.32 GPa and 6.16 GPa *(close to maximum $T_c$ ~ 27 K)* PQed at $T_Q$ = 4.2 K and 77 K, respectively; and in Fig. 4b for 9.65 GPa *(non-SC)* PQed at $T_Q$ = 77 K. As is evident from the $R(T,P_A)$s/$R(50 K,P_A)$ in Fig. 4a, $P_A$ ~ 6 GPa has induced a SC state in the non-SC Cu-doped FeSe with a $T_c$ ~ 26 K (navy and blue); this SC state has been PQed at $P_Q$ = 6.16 GPa and $T_Q$ = 4.2 K (red) and at $P_Q$ = 6.32 GPa and $T_Q$ = 77 K (green), respectively. Disappearance of the SC phase after thermal cycling up to 300 K (Fig. 4a, orange and brown) demonstrates that the SC states induced by $P_A$ ~ 6 GPa have been retained at ambient with $T_c$ ~ 26 K via PQ at 4.2 K and 77 K, respectively. As shown in Fig. 4b, $P_A$= 9.65 GPa turns the sample to an insulating state (blue); upon PQ at $T_Q$ = 77 K it becomes insulating (green); and the sample remains in the non-SC state after thermal cycling to 300 K (orange), suggesting that the insulating state PQed at 9.65 GPa and 77 K is stable up to 300 K. The thermal stability ranges of the PQed SC states at $P_Q$ = 6.08 GPa and $T_Q$ = 4.2 K and at $P_Q$ = 5.95 GPa and $T_Q$ = 77 K are shown in Figs. 4c and 4d, respectively. They show that the state PQed at a lower $T_Q$ possesses a wider thermal stability range. The anomalies observed in R(T) upon warming right after PQ (Fig. 4e) correlate qualitatively with the thermal stability of the PQed phases (Figs. 4c-d). Figure 4f demonstrates that the PQed SC phase at $P_Q$ = 6.67 GPa and $T_Q$ = 77 K remains unchanged for at least 7 days after thermal cycling between 50 K and 4.2 K. All PQed $T_c$s of Cu-doped FeSe are summarized in Fig. 3. Unlike in their pristine unpressured state, the two different Cu-doped FeSe samples both behave similarly to FeSe under pressures, but with their phase boundaries shifted to higher values, as displayed in Figs. 1 and 3 and Extended Data Fig. 4, due to the Cu-doping effect. While PQ works for both pure and Cu-doped FeSe in retaining at ambient the pressure-enhanced or -induced SC states, the effect of $T_Q$ on the $T_c$ of the PQed SC phase for Cu-doped FeSe is smaller than that for FeSe, due to the possible change in the electronic structure resulting from doping. This suggests that doping can help adjust the quenching parameters.

To gain a better understanding of the PQ effects on FeSe, we performed *ab initio* simulations to evaluate the phase transition energy barriers between different phases via SSNEB[6]. As shown in Extended Data Fig. 5a-b, the phase transition energy barrier between the orthorhombic[22] and tetragonal[23] phases is small. For instance, the energy barrier is 3 meV/atom at 6 GPa, which is lower than the energy barrier of 6 meV/atom at 0 GPa, suggesting that the transition temperature between these two phases at HP should be lower than that at ambient, in agreement with the experimental observations. Nevertheless, the small energy barrier between those two structures ensures that FeSe could preserve the structure phase from one transfer to the other when PQed from above 2 GPa to ambient at low temperatures, as well as the superconductivity. On the other hand, the phase transition energy barrier from the hexagonal[24] to the tetragonal phase is significantly larger, about 0.189 and 0.193 eV/atom at 8 and 11 GPa, respectively (Extended Data Fig. 5c-d). We also noticed that the tetragonal phase is energetically more favorable than the hexagonal phase at simulated pressures. The phase transition between the tetragonal and hexagonal phases will occur at 15 GPa based on our simulation, in agreement with previous calculations[25]. The estimated energy barriers are comparable to that between graphite and cubic diamond, around 0.21 eV/atom at 10 GPa[26], suggesting that the hexagonal phase could be preserved during the quenching process once it is formed, as we observed in our experiments. The energy barrier is high enough to prevent FeSe returning to the orthorhombic phase from the hexagonal phase at ambient pressure and 300 K, which is consistent with our experiments at $P_Q$ = 11.12 GPa and $T_Q$ = 77 K shown in Fig. 2e.

In conclusion, we have demonstrated that the pressure-induced/enhanced superconducting phases with high $T_c$ and the pressure-induced semiconducting phases in FeSe and Cu-doped FeSe can be stabilized at ambient without pressure by pressure-quenching at chosen pressures and temperatures. Theses pressure-quenched phases have been shown to be stable at up to 300 K and for up to at least 7 days. The observations suggest that the recently reported room-temperature superconductivity in hydrides close to 300 GPa may be retained without pressure, making possible the ubiquitous applications of superconductivity envisioned.

**Figures and legends**

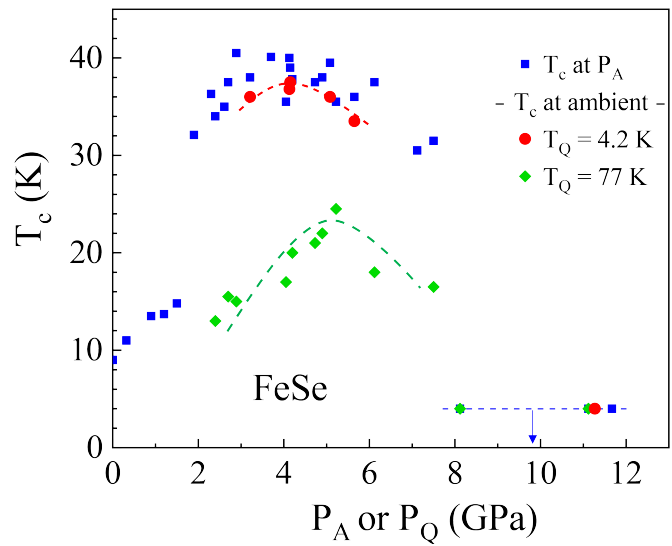

**Fig. 1 | $T_c$ as a function of $P_A$ or $P_Q$ for single-crystalline FeSe.** High-pressure $T_c$ ($P_A$) at $P_A$ (blue squares), and $T_c$ ($P_Q$) at ambient for FeSe PQed at $P_Q$ and $T_Q$ = 4.2 K (red circles) and $T_Q$ = 77 K (green diamonds), respectively.

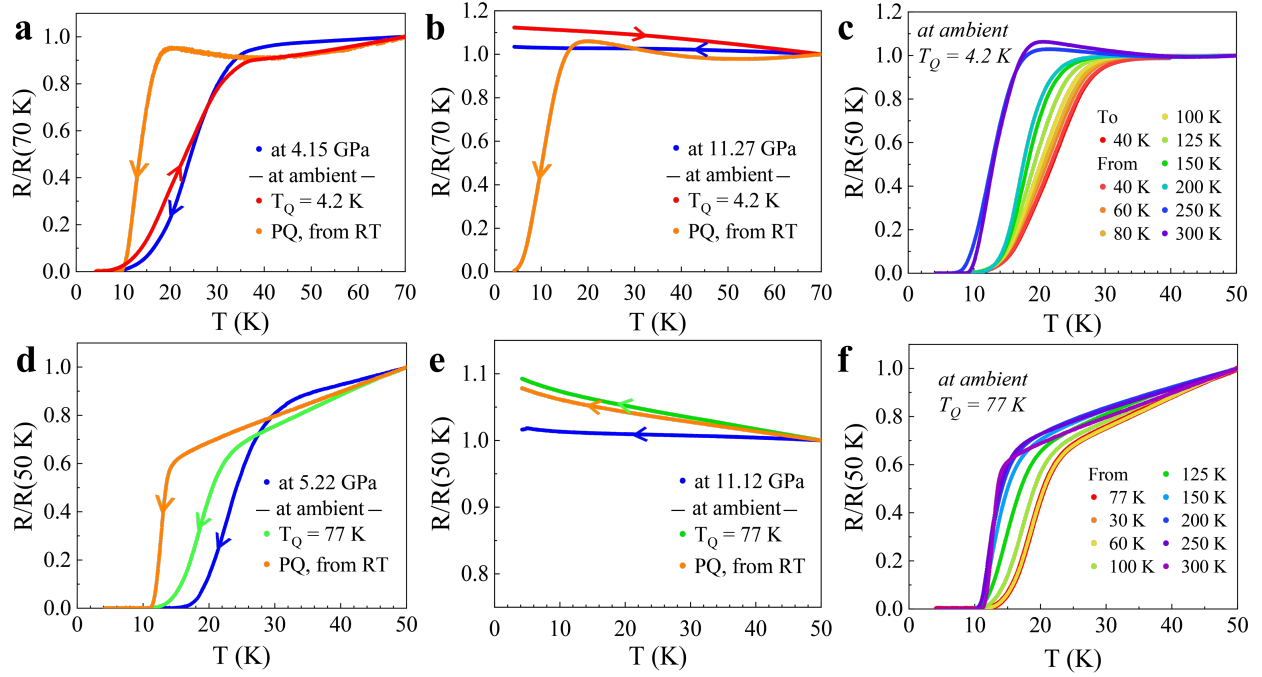

**Fig. 2 | Pressure-quenching the single-crystalline FeSe**. R(T)/R(70 K) or R(T)/R(50 K) under $P_A$ and at ambient after PQ: **a.** at $P_A$ = 4.15 GPa (blue), and at ambient after PQ at 4.15 GPa and 4.2 K on warming (red) and on cooling after warming to 300 K (orange); **b.** at $P_A$= 11.27 GPa (blue), and at ambient after PQ at 11.27 GPa and 4.2 K on warming (red) and on cooling after warming to 300 K (orange); **c.** after PQ at 4.13 GPa and 4.2 K, sequential thermal cycling up to 300 K; **d.** at $P_A$ = 5.22 GPa (blue), and at ambient after PQ at 5.22 GPa and 77 K on cooling (green) and on cooling after warming to 300 K (orange); **e.** at $P_A$ = 11.12 GPa (blue), and at ambient after PQ at 11.12 GPa and 77 K on cooling (green) and on cooling after warming to 300 K (orange); and **f.** after PQ at 5.22 GPa and 77 K, sequential thermal cycling up to 300 K.

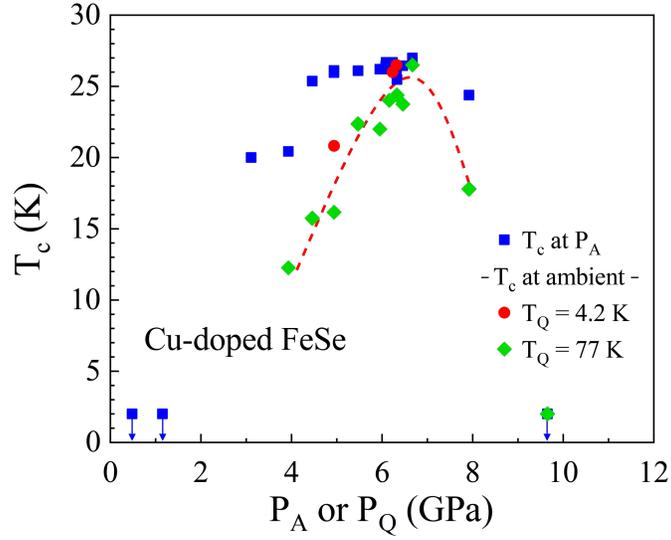

**Fig. 3 | $T_c$ as a function of $P_A$ or $P_Q$ for single-crystalline Cu-doped FeSe.** $T_c$ ($P_A$) at $P_A$ (blue squares); and at $T_c$ ($P_Q$) at ambient for the samples PQed at $P_Q$ and $T_Q$ = 4.2 K (red circles) and at $T_Q$ = 77 K (green diamonds), respectively.

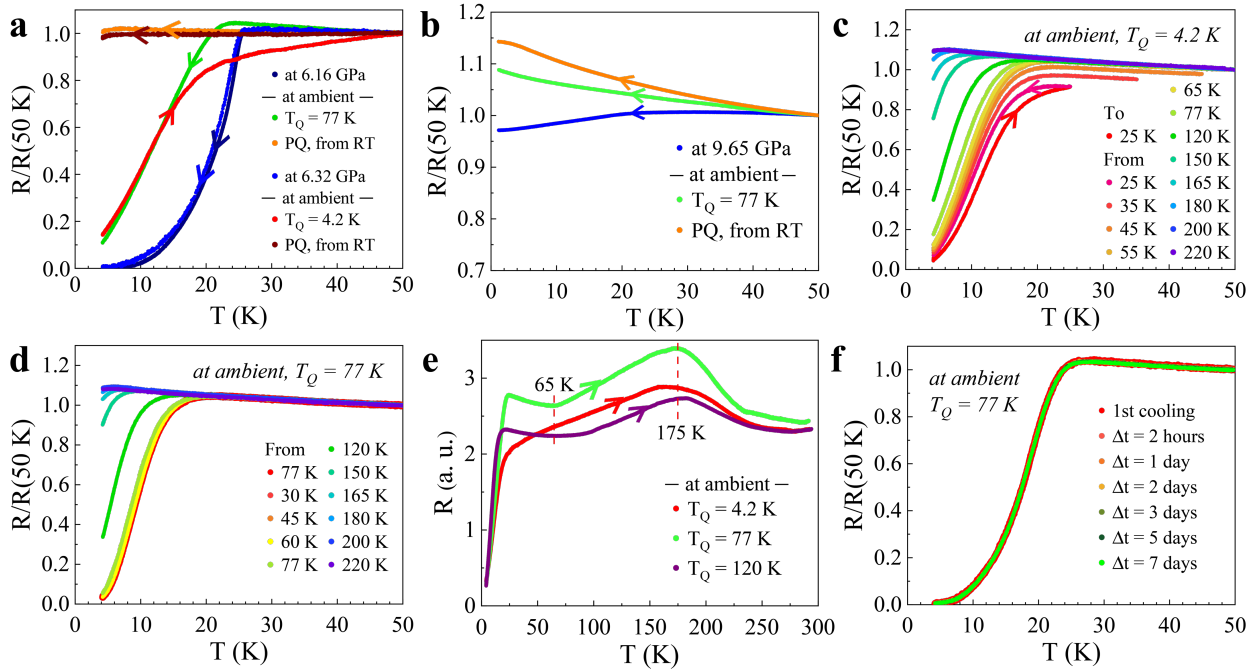

**Fig. 4 | Pressure-quenching the single-crystalline Cu-doped FeSe.** R(T)/R(50 K) under $P_A$ and at ambient after PQ, and testing the stability of the PQed phases: **a.** at $P_A$ = 6.16 GPa (navy) and

6.32 GPa (blue), and at ambient after PQ at 6.16 GPa and 77 K (green), and at 6.32 GPa and 4.2 K (red), and on cooling after warming to 300 K (orange and brown); **b.** at $P_A$= 9.65 GPa (blue), at ambient after PQ at 9.65 GPa and 77 K (green), and on cooling after warming to 300 K (orange); **c.** at ambient after PQ at 6.08 GPa and 4.2 K, sequential thermal cycling up to 220 K; **d.** at ambient after PQ at 5.95 GPa and 77 K, sequential thermal cycling up to 220 K; **e.** R(T) at ambient for the same sample subjected to different PQ conditions: $P_Q$ = 6.31 GPa and $T_Q$ = 4.2 K (red), $P_Q$ = 6.16 GPa and $T_Q$ = 77 K (green), and $P_Q$ = 6.51 GPa and $T_Q$ = 120 K (purple); and **f.** repeated thermal cycling at ambient from 50 K for the sample PQed at 6.67 GPa and 77 K.

**Methods**

Single crystals of $Fe_{1.01-x}Cu_xSe$ (x =0, 0.03, and 0.035) were grown using the chemical vapor transport (CVT) method[27]. Stoichiometric high-purity Fe, Cu, and Se powders were thoroughly mixed and loaded into a quartz tube. $AlCl_3$ and KCl powders were added as the transport agents. After the evacuated quartz tube was sealed, it was placed into a two-zone tube furnace, in which the temperatures of the hot and cold positions were maintained at 420 °C and 330 °C, respectively. After 20 days, single crystals with an average size of 3 × 3 × 0.1 $mm^3$ were grown around the region of the quartz tube's cold zone. Chemical composition was determined by energy-dispersive spectroscopy (EDS) using a Tescan Lyra scanning electron microscope (SEM) equipped with an EDS detector (Oxford Instruments).

For resistivity measurements conducted in this investigation, pressure was applied to the samples using a symmetric-type diamond anvil cell with a culet size of 500 μm. The stainless-steel gasket was insulated with Stycast 2850. The sample's chamber diameter is 250 μm, where either sodium chloride or cubic boron nitride is used as the pressure medium. Samples were cut into thin squares with a diagonal of ~ 180 μm and thickness of ~ 10 μm. The pressure was determined using the ruby fluorescence scale or the diamond Raman scale at room temperature. The samples' contacts were arranged in a Van der Pauw configuration and data were collected using a Keithley 2182A/6221 low-resistance measurement setup. Measurements were conducted in a homemade cooling system that can be cooled to 1.2 K by pumping on the liquid-helium space. Pressure-quenching was performed by releasing the screws at target temperatures with a small residual pressure $P_R$ < 0.2 GPa to maintain the electrical connectivity for resistivity measurements, and the $P_R$ was measured at room temperature.

Our calculations were performed within the framework of density functional theory via the generalized gradient approximation GGA + U method implemented in the Vienna ab initio simulation package (VASP)[28]. The electron-ion interactions were represented by means of the all-electron projector augmented wave (PAW) method[29], where $3d^64s^2$ and $4s^24p^4$ are treated as the valence electrons for Fe and Se, respectively. We used the Dudarev implementation[30] with on-site coulomb interaction U = 5.0 eV and on-site exchange interaction J = 0.8 eV[31] to treat the localized 3d electron states. The Perdew-Burke-Ernzerhof (PBE) functional in the generalized gradient approximation (GGA) was used to describe the exchange-correlation potential[32,33]. The plane-wave energy cutoff of 400 eV and a dense k-point grid of spacing $2\pi \times 0.03$ Å$^{-1}$ in the Monkhorst-Pack scheme were used to sample the Brillouin zone. Structural relaxations were performed with forces converged to less than 0.05 eV Å$^{-1}$. To determine the energy barriers, we used the solid-state nudged elastic band method (SSNEB)[6] implemented in VASP. The NEB path was first constructed by linear interpolation of the atomic coordinates and then relaxed until the forces on all atoms were < 0.05 eV/Å. Seven images were simulated between the initial and final states.

**Methods References**

**Author contributions**

data. B.G., T.C., and P.D. grew the samples. Y.X., X.L., K.Y., Z.W., and Y.M. provided theoretical interpretation. C.-W.C., L.D., and Z.W. provided overall supervision.

**Acknowledgments**

We thank Prof. L. L. Sun, C. Huang, and J. Guo at IoP, CAS, for efforts on the synchrotron X-ray experiments under high pressure at beamline 4W2 of the Beijing Synchrotron Radiation Facilities. The work performed at the Texas Center of Superconductivity at the University of Houston is supported by US Air Force Office of Scientific Research Grant FA9550-15-1-0236 and FA9550-20-1-0068, the T. L. L. Temple Foundation, the John J. and Rebecca Moores Endowment, and the State of Texas through the Texas Center for Superconductivity at the University of Houston. The FeSe and Cu-doped FeSe single crystal growth work at Rice University is supported by the US Department of Energy, Basic Energy Sciences, under Contract No. DE-SC0012311 (P.D.).

**Competing interest declaration**

The authors declare no competing interests.

**Data availability**

All data that support the findings of this study are available from the corresponding authors on request.

**Extended Data**

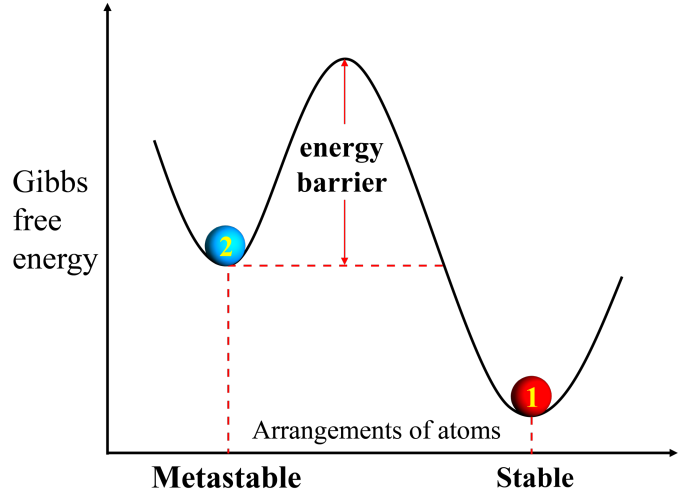

**Extended Data Fig. 1 | Schematic diagram of Gibbs free energy and the energy barrier between the metastable and stable states.**

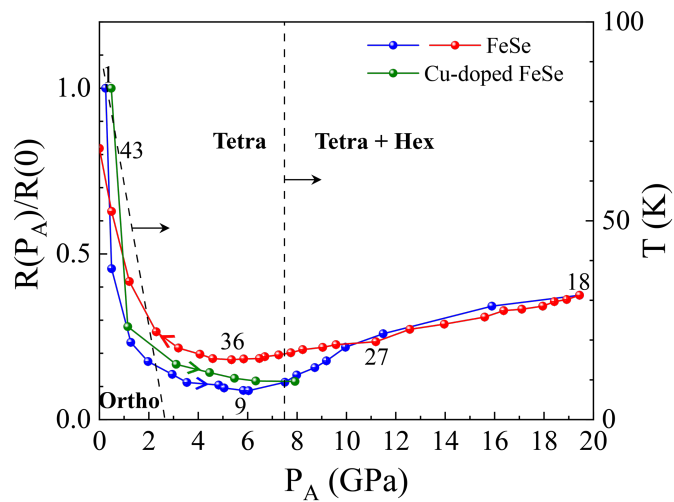

**Extended Data Fig. 2 | $R(P_A)/R(0)$ of FeSe and Cu-doped FeSe single crystals at 300 K during pressure cycling.** Dashed lines represent phase boundaries for FeSe[16,20]. Numbers denote the sequential order of the experimental runs.

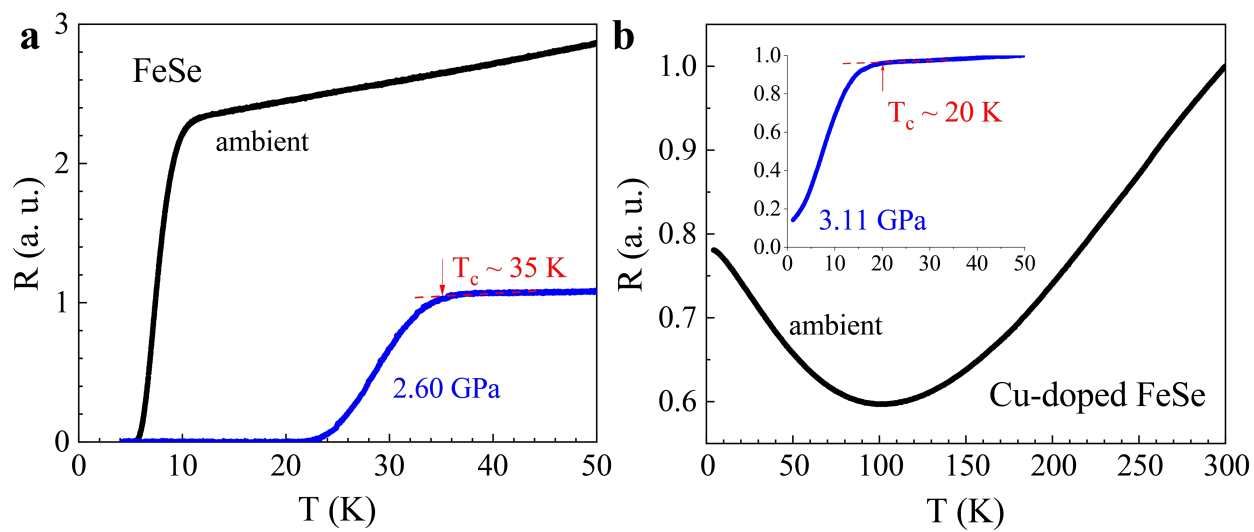

**Extended Data Fig. 3 | Resistance as a function of temperature for FeSe and Cu-doped FeSe single crystals. a,** FeSe at ambient (black) and at 2.60 GPa (blue). **b,** Cu-doped FeSe at ambient and (inset) at 3.11 GPa. Red dashed lines and arrows define the $T_c$.

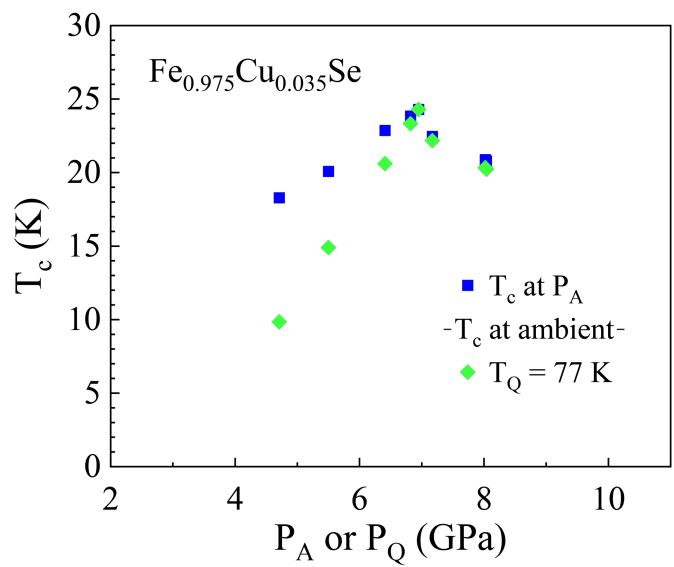

**Extended Data Fig. 4 | $T_c$ as a function of $P_A$ or $P_Q$ for single-crystalline Cu-doped FeSe.** High-pressure $T_c$ ($P_A$) at $P_A$ (blue squares); and $T_c$ ($P_Q$) at ambient for Cu-doped FeSe sample PQed at $P_Q$ and $T_Q = 77$ K (green diamonds).

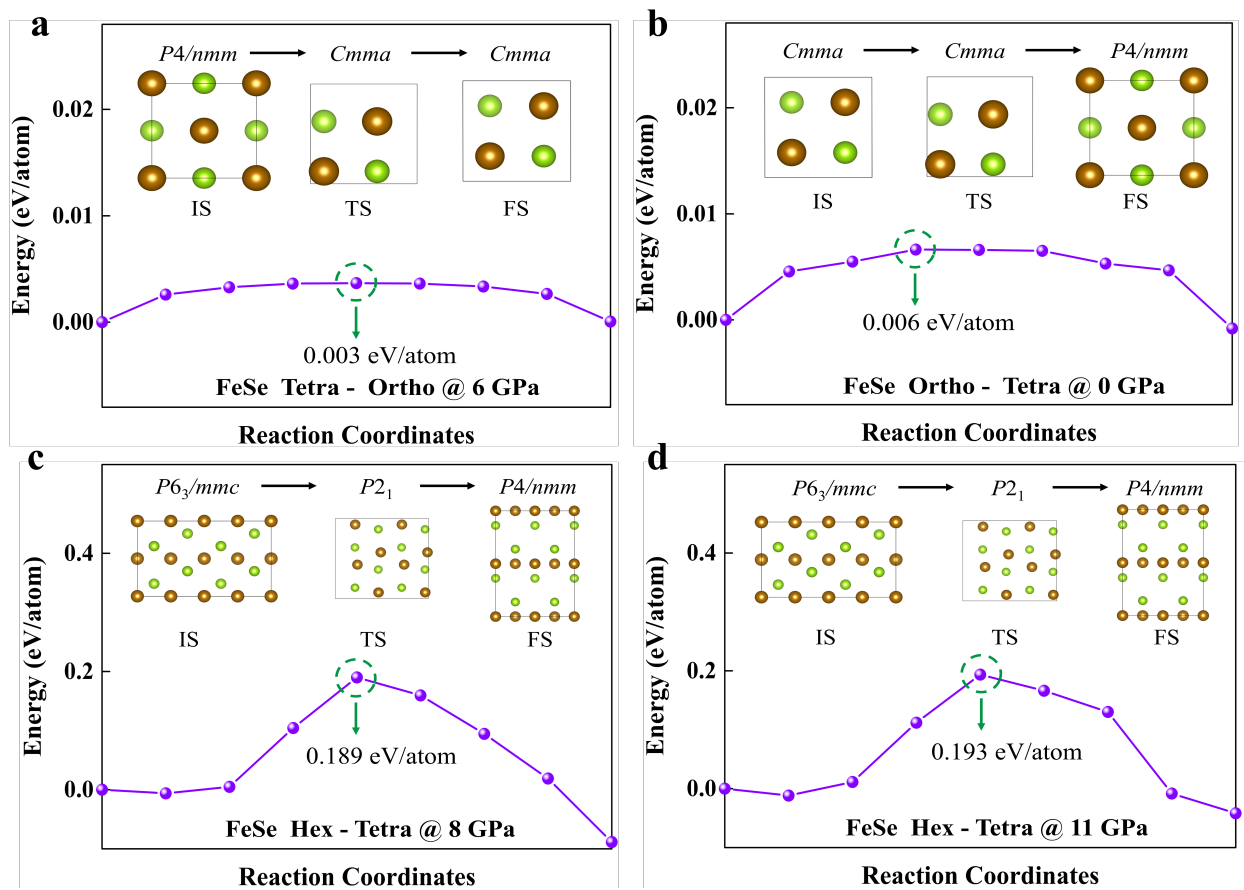

**Extended Data Fig. 5 | Energy barrier between different phases of FeSe.** Calculated energy barrier from **a,** the tetragonal phase to the orthorhombic phase at 6 GPa and **b,** the orthorhombic phase to the tetragonal phase at 0 GPa. Calculated energy barrier from the hexagonal phase to the tetragonal phase at **c,** 8 GPa and **d,** 11 GPa. Insets show the side views of corresponding structures including the initial state (IS), the transition state (TS), and the final state (FS) along the c axis. The energy barrier was calculated through the solid-state nudged elastic band method in which seven images were used. The arrows show the transition state that is the image with the highest energy and the estimated energy barrier. The green and brown spheres represent elemental Se and Fe, respectively.